\documentclass[letter,12pt]{article}
\pdfoutput=1
\usepackage{jheppub}
\usepackage{hyperref}
\usepackage{graphicx,epstopdf,amsmath,amsfonts,amssymb,appendix,comment} 
\usepackage{color,slashed,setspace,footnote,multirow,longtable,braket}
\usepackage{subcaption}
\usepackage{epsfig,mathrsfs,latexsym,color,url,etoolbox}
\usepackage{bbm}
\usepackage[normalem]{ulem}
\definecolor{nicered}{rgb}{0.7,0.1,0.1}
\definecolor{nicegreen}{rgb}{0.1,0.5,0.1}
\definecolor{CarnationPink}{rgb}{1.0, 0.65, 0.79}
\DeclareMathAlphabet{\mathbbold}{U}{bbold}{m}{n}    

\usepackage{comment}
\usepackage{mathabx}

\usepackage{float}

\allowdisplaybreaks

\usepackage{dcolumn}
\usepackage{bm}
\usepackage[table]{xcolor}
\usepackage{multirow}
\usepackage{comment}
\usepackage{url}
\usepackage{ulem}
\usepackage{tcolorbox}
\usepackage[T1]{fontenc}
\usepackage[compat=1.1.0]{tikz-feynman}

\definecolor{kjkblue}{rgb}{0.39, 0.589, 0.6914}

\newcommand\ddfrac[2]{\frac{\displaystyle #1}{\displaystyle #2}}

\DeclareMathAlphabet{\mathpzc}{OT1}{pzc}{m}{it}

\newcommand{\Ehad}{{E_{\rm had}}}

\def\Fermilab{Theoretical Physics Department, Fermilab, P.O. Box 500, Batavia, IL 60510, USA}
\def\UChicago{Department of Physics, University of Chicago, Chicago, IL 60637, USA}
\def\UCIrvine{Department of Physics and Astronomy, University of California, Irvine, CA 92697, USA}

\begin{document}

\preprint{FERMILAB-PUB-22-726-T}

\title{The impact of neutrino-nucleus interaction modeling on new physics searches}

\author[1]{Nina M. Coyle}
\author[2,3]{Shirley Weishi Li}
\author[2]{Pedro A. N. Machado}
\affiliation[1]{\UChicago}
\affiliation[2]{\Fermilab}
\affiliation[3]{\UCIrvine}
\emailAdd{ninac@uchicago.edu}
\emailAdd{shirleyl@fnal.gov}
\emailAdd{pmachado@fnal.gov}

\date{\today}

\abstract{
Accurate neutrino-nucleus interaction modeling is an essential requirement for the success of the accelerator-based neutrino program.
As no satisfactory description of cross sections exists, experiments tune neutrino-nucleus interactions to data to mitigate mis-modeling. 
In this work, we study how the interplay between \emph{near detector tuning} and cross section mis-modeling affects new physics searches.
We perform a realistic simulation of neutrino events and closely follow NOvA's tuning, the first published of such procedures in a neutrino experiment.
We analyze two illustrative new physics scenarios, sterile neutrinos and light neutrinophilic scalars, presenting the relevant experimental signatures and the sensitivity regions with and without tuning.
While the tuning does not wash out sterile neutrino oscillation patterns, cross section mis-modeling can bias the experimental sensitivity. In the case of light neutrinophilic scalars, variations in cross section models completely dominate the sensitivity regardless of any tuning.
Our findings reveal the critical need to improve our theoretical understanding of neutrino-nucleus interactions, and to estimate the impact of tuning on new physics searches. 
We urge neutrino experiments to follow NOvA's example and publish the details of their tuning procedure, and to develop strategies to more robustly account for cross section uncertainties, which will expand the scope of their physics program.
}

\maketitle
\flushbottom

\clearpage

\section{Introduction}
\label{sec:Introduction}

The neutrino sector is the least understood of the Standard Model (SM).
Several questions about neutrinos, of both experimental and theoretical character, remain open to this date, including the conservation of lepton number, the mass mechanism, the mass ordering, leptonic charge-parity violation, and puzzling experimental results which remain to be understood, among others.
One of the reasons some of these questions remain unanswered lies in neutrinos' small, weak-only interactions.
Studying these fermions requires special experimental setups, with intense sources and typically huge detectors.
As a consequence, the neutrino sector is widely regarded as a promising portal to beyond the standard model (BSM) physics, while neutrino experiments are excellent tools to search for novel BSM signatures.

Of particular interest are accelerator neutrino experiments, in which an intense neutrino beam is produced by impinging protons on a target, and detectors are placed near and/or far from the neutrino beam origin.
While such a setup provides a rich environment to study standard and beyond standard physics, it also comes with challenges.
These experiments measure neutrinos through their interactions on nuclei, which are quite complex and not fully understood processes. 
In the energy range between a few hundred MeV to a few GeV, the kinematics span perturbative and non-perturbative regimes, with the hadronic degrees of freedom being a mixture of nucleons, resonances, and partons~\cite{Kronfeld:2019nfb, Mosel:2019vhx, Alvarez-Ruso:2020ezu}. 
In addition, nuclear effects can play an important role in the final states of neutrino-nuclei interactions~\cite{Bertini:1963zzc, Cugnon:1980zz, Bertsch:1984gb, Bertsch:1988ik, Cugnon:1996xf, Boudard:2002yn, Sawada:2012hk, Uozumi:2012fm, Golan:2012wx, Rocco:2020jlx, Isaacson:2020wlx, Dytman:2021ohr}. 
Although there have been many recent studies on understanding neutrino-nucleus scattering cross sections~\cite{Meyer:2016oeg, MINERvA:2019rhx, Ankowski:2019mfd, MINERvA:2020anu, MINERvA:2020zzv, Borah:2020gte, T2K:2020jav, T2K:2021naz, Tomalak:2021qrg, MINERvA:2021owq, NOvA:2021eqi}, we still do not have a sound theoretical framework to compute the cross sections or estimate their uncertainties~\cite{Ankowski:2020qbe, electronsforneutrinos:2020tbf, CLAS:2021neh}.
On top of that, the neutrino flux is driven by meson production and decay, which suffer from considerable QCD hadronization uncertainties~\cite{MiniBooNE:2008hfu, T2K:2015sqm, MINERvA:2016iqn}.

A data-driven approach is typically sought to mitigate the large uncertainties associated with the neutrino flux and interaction models. 
Commonly, neutrino event spectra at near detectors (ND) are used to calibrate the neutrino flux and interaction cross section models.
We will refer to this procedure, which will be explained in detail later, as ND tuning. 
This information is then used in the far detector (FD) to extract oscillation parameters and other neutrino properties.
Although the primary role of NDs is to provide these inputs for physics searches at the FDs, NDs themselves are very special.
Due to the proximity from the neutrino source and large size, they will typically have a very high neutrino event rate, accruing millions of events over the lifetime of an experiment, as opposed to FD statistics of $\mathcal{O}(1000)$ events or less.
Such large statistical data sets have an enormous potential to probe new physics in the neutrino sector, particularly for models which may affect the ND more than the FD~\cite{Batell:2009di, deNiverville:2011it,  deNiverville:2012ij, Dharmapalan:2012xp,Izaguirre:2013uxa, Batell:2014yra, Dobrescu:2014ita, Coloma:2015pih, deNiverville:2016rqh, Magill:2016hgc, MINOS:2017cae, Bertuzzo:2018itn, Falkowski:2018dmy, Magill:2018tbb, Ballett:2018ynz, deGouvea:2018cfv, Ballett:2018uuc, deNiverville:2018dbu, Jordan:2018gcd, Jordan:2018qiy, Ballett:2019xoj, Harnik:2019zee, ArgoNeuT:2019ckq, Altmannshofer:2019zhy, Tsai:2019mtm, Batell:2019nwo, Arguelles:2019xgp, DeRomeri:2019kic, deGouvea:2019wav, Berryman:2019dme, DUNE:2020fgq}.
Because we do not have the perfect model of neutrino-nucleus interactions, and the cross section and flux modeling is tuned with the ND data, an obvious question is raised: \emph{how much does the interplay between ND tuning and cross section mis-modeling interfere with new physics searches?}

This paper aims to study the interplay between cross section mis-modeling and searches for BSM physics.
To estimate the impact on BSM searches, we analyze a couple of illustrative scenarios, presenting their signatures before and after tuning.
Because the tuning procedure is not uniquely defined and its impact depends on the technical details, we follow the procedure described in a recent, detailed publication of the NOvA collaboration~\cite{NOvA:2020rbg} as closely as possible. 
We choose to follow NOvA's approach because they are the only collaboration that published the details of their tuning with codes at the time of this work; we strongly encourage other experiments to follow NOvA's example and publish their tunings.
For this work, we generate neutrino events with state-of-the-art tools and apply NOvA's tuning on an event-by-event basis.
We identify and discuss how the smoking gun signatures of new physics models are affected.
While some signatures, such as sterile neutrino oscillations, are robust against tuning, mis-modeling of neutrino interactions can bias the experimental sensitivity.
On the other hand, the sensitivity to other types of BSM scenarios can be overwhelmed by cross section mis-modeling.

This is the first quantitative study exploring the impact of tuning on new physics searches. We identify conceptual lessons that should be applicable to other tuning procedures, detector set ups, analysis details, and BSM scenarios. 
For instance, we expect our findings to hold to some extent even in BSM searches that do not employ a ND tuning but instead rely on a simultaneous fit to near and far detector data (see, e.g., Ref.~\cite{MINOS:2017cae}). 
Our goal is not to faithfully reproduce any experimental analyses.
We hope our results will encourage theorists and experimentalists to consider the impact of ND tuning on new physics scenarios and to carefully estimate systematic uncertainties related to neutrino fluxes and cross sections.

\section{Near detector tuning}
\label{sec:Methods}

Near detectors in long-baseline neutrino oscillation experiment are designed to mitigate the uncertainties from both the neutrino beams and neutrino-nucleus cross sections. Conceptually, this procedure may seem straightforward:
\begin{equation}
P(\nu_\alpha\rightarrow\nu_\beta; E_\nu) = C\ddfrac{
\frac{dN^\mathrm{FD}_\beta}{dE_\nu}\bigg/\sigma_\beta(E_\nu)}{\frac{dN^\mathrm{ND}_\alpha}{dE_\nu}\bigg/\sigma_\alpha(E_\nu)} , 
\end{equation}
where $P(\nu_\alpha\rightarrow\nu_\beta;E_{\nu})$ is the oscillation probability the experiment is trying to measure, $C$ is a constant accounting for detector sizes and distances to the source, the superscripts ND and FD denote near and far detectors, and $dN_{\alpha,\beta}/dE_{\nu}$ are the event rates of $\nu_{\alpha,\beta}$ in terms of true neutrino energy. 
The differences between $\sigma_\alpha$ and $\sigma_\beta$, if $\alpha \neq \beta$, could in principle be computed theoretically~\cite{Tomalak:2021qrg}. 
The oscillation probability can be directly derived from the unoscillated event rate $dN_\alpha/dE_\nu$ measured at the ND and the oscillated event rate $dN_\beta/dE_\nu$ measured at the FD. 

In reality, this ``logical division'' is not practical because one cannot directly measure neutrino energy, the near and far detectors have different systematics, and even the unoscillated neutrino fluxes are different at the near and far detectors solely due to the different solid angles. 
We can appreciate this more concretely by studying the following equation,
\begin{equation}
\label{eq:prob}
\frac{N_{\rm FD}}{N_{\rm ND}} (E_\mathrm{reco})= C \ddfrac{\int dE_\nu \frac{d\phi^{\rm FD}_\alpha}{dE_\nu}P(\nu_\alpha\to\nu_\beta; E_\nu) \sigma_\beta(E_\nu) \mathcal{M}^\mathrm{FD}_\beta(E_\nu, E_\mathrm{reco}) }
{\int dE_\nu \frac{d\phi^{\rm ND}_\alpha}{dE_\nu}\sigma_\alpha(E_\nu)\mathcal{M}^\mathrm{ND}_\alpha(E_\nu, E_\mathrm{reco})} ,
\end{equation}
where $N_{\rm FD/ND}$ encode the reconstructed neutrino event spectra, $d\phi^\mathrm{FD/ND}_\alpha/dE_\nu$ are the fluxes at the far and near detectors without oscillation, $\sigma_{\alpha,\beta}(E_\nu)$ are the total cross sections, and $\mathcal{M}^\mathrm{FD/ND}_\alpha$ are the migration matrices. 
The challenge here is that experiments measure the left-hand side of Eq.~\eqref{eq:prob} and they need to infer the oscillation probability $P$ on the right-hand side.
The main difficulty is encoded in the term $\sigma_\alpha(E_\nu)\mathcal{M}_\alpha(E_{\nu},E_{\mathrm{reco}})$, where the reconstruction of the true neutrino energy depends on the details of neutrino-nucleus interaction, as well as detector responses to different final-state particles. An obvious example is neutrons, for which both the modeling of neutrons produced by neutrino interactions and the corresponding detector responses are relevant. 
Currently, predictions on the number of outgoing nucleons in a neutrino-nucleus scattering event, as well as their energy and isospin, differ drastically among generators~\cite{Ankowski:2019mfd,CLAS:2021neh}. In addition, neutron detector responses suffer from significant uncertainties due to neutron propagation and event reconstruction~\cite{Friedland:2018vry, CAPTAIN:2019fxo, Friedland:2020cdp}.

Because of these complications, oscillation experiments adopt a near-detector tuning procedure. Assuming SM physics, one can predict the measured neutrino energy spectrum in the ND with an accelerator neutrino beam simulation, an event generator such as \texttt{GENIE}~\cite{Andreopoulos:2009rq}, \texttt{NuWro}~\cite{Golan:2012rfa}, \texttt{GiBUU}~\cite{Buss:2011mx}, or \texttt{ACHILLES}~\cite{Isaacson:2022cwh} that simulates neutrino-nucleus interaction cross sections and final states, and a detector simulation predicting the migration matrix. When the predicted and measured neutrino spectra disagree, one can modify the cross section simulations until they match.
Because of the nature of such calculations and the complexities of these simulation packages, there is no unique, agreed-upon way to tune the models. 
One can vary the model parameters, adopt alternative models, or take a model-agnostic approach and add more degrees of freedom. 

In addition, it is well appreciated that neutrino beams have sizable uncertainties and they interfere with cross section uncertainties. Different experiments also treat the flux tuning and the cross section tuning differently. NOvA, for instance, only uses hadronic production data, in-situ measurements of horn position and current, beam parameters, etc., and MINERvA neutrino-electron scattering data to tune their flux prediction~\cite{NOvA:2021nfi}. T2K, on the other hand, uses ND neutrino-nucleus scattering data in addition to auxiliary data to tune their flux and cross section models at the same time~\cite{T2K:2021xwb}.

In this work, we follow the tuning procedure outlined by NOvA~\cite{NOvA:2018gge}. The NOvA experiment is a long-baseline experiment comprised of two scintillator detectors (CH$_2$) placed along a $\nu_{\mu} / \bar{\nu}_{\mu}$ beamline produced by the NuMI facility at Fermilab: 
a near and a far detector placed 1~km and 810~km from the beam source, respectively.
The 14~kton FD observes the muon neutrino beam after long-baseline oscillations, intended to measure oscillation parameters including $\Delta m_{32}^2$ and $\theta_{23}$. The smaller 0.3~kton ND is nearly identical to the FD to minimize systematic uncertainties.

The ND tune procedure is detailed in Ref.~\cite{NOvA:2020rbg}. Neutrino interactions with the material in the ND are first generated using \texttt{GENIE} v2.12.2 with the default models and parameters. Then, the following changes are applied to the \texttt{GENIE} simulation based on auxiliary theoretical and experimental studies, i.e., {\it NOvA ND data is not used for this step}: 
\begin{itemize}
	\item adjusting the value of axial mass, $m_A$, from 0.99 to 1.04~GeV, based on recent re-analysis~\cite{Meyer:2016oeg} of neutrino-deuterium scattering data;
	\item modifying the momentum distributions of the initial nucleons for quasi-elastic scattering, based on a MINERvA study~\cite{Gran:2017psn};
	\item lowering the magnitude of neutrino-scattering in non-resonance pion production regime by 57\%, motivated by re-analysis of old bubble chamber data~\cite{Rodrigues:2016xjj};
	\item suppressing delta resonance production in low-$Q^2$ region, motivated by measurements by MiniBooNE~\cite{MiniBooNE:2010eis}, MINOS~\cite{MINOS:2014axb}, MINERvA~\cite{MINERvA:2016sfc,MINERvA:2017okh}, and T2K~\cite{T2K:2019yqu}. 
\end{itemize}
All these changes are to improve the baseline model in \texttt{GENIE}.

\begin{figure}[t] 
	\centering
	\includegraphics[width=\textwidth]{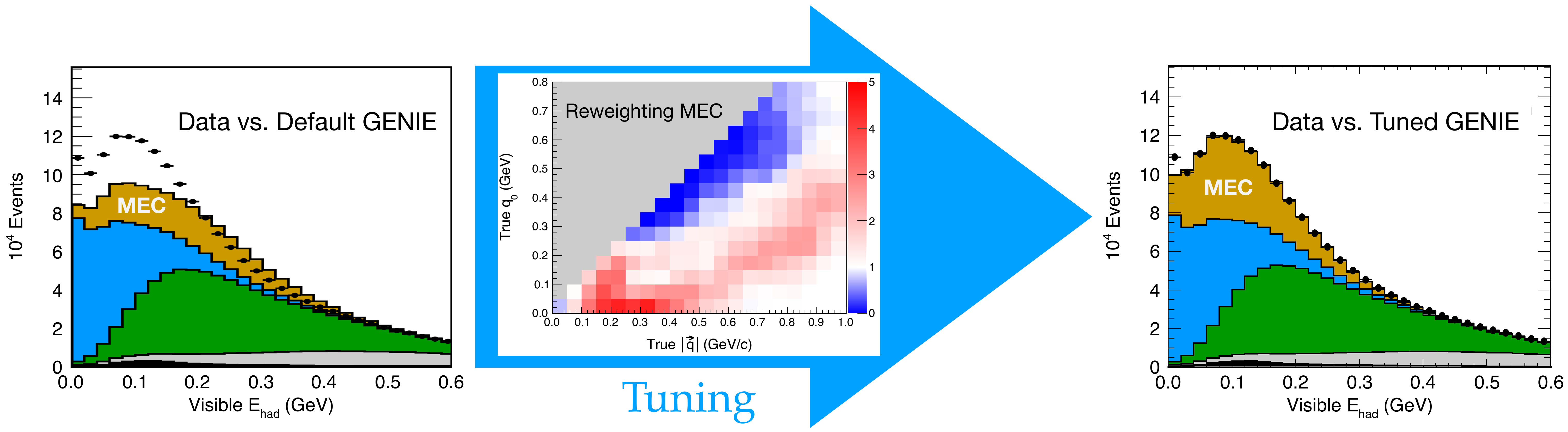}
	\caption{Schematic illustration of NOvA's tuning. On the left, we show the NOvA data versus the output of default \texttt{GENIE} 2.12.2 (colored histograms). The tuning happens in two steps, first, a few adjustments to \texttt{GENIE} are made, such as redefining $m_A$, and then the meson-exchange current (MEC) component of the cross section is reweighted to data (see text for details). The result is the plot on the right, where the output of tuned \texttt{GENIE} matches the data well. Details have been omitted for clarity. These figures are adapted from Ref.~\cite{NOvA:2020rbg}.
	\label{Fig:flowchart}
	}
\end{figure}

After these changes are applied to \texttt{GENIE}, there are still large discrepancies between the measured neutrino spectrum in NOvA ND and the simulated spectrum. The last step of NOvA's ND tune is the crucial step of which we are studying the effect. To understand how it works, let us first define the kinematics of neutrino interactions, then explain how kinematic variables are measured in NOvA. 
Theoretically, an incoming neutrino with energy $E_\nu$ produces an outgoing lepton with energy $E_l$ and a hadronic system. The four-momentum transfer is $(q_0, \vec{q})$ with $q_0 = E_\nu - E_l$. 
In NOvA, as in any experiment, the neutrino energy is not directly observable and is instead measured through $E_\nu^\mathrm{reco} \equiv E_l^\mathrm{reco} + q_0^\mathrm{reco}$. For charged-current interactions, which are the signal channels for oscillation analyses, $E_l$ can be measured relatively well either from the total scintillation light associated with the electron or muon or from the muon range, i.e., $E_l^\mathrm{reco} \simeq E_l$.   
The transferred energy, $q_0$, gets distributed between kinetic energy for knocked out nucleons, total energy for mesons, and binding energy of the initial nucleus. 
The sum of nucleon kinetic energy, $T$, and meson total energy $E$, roughly proportional to the amount of scintillation light detected, is called the hadronic energy, $\Ehad$, i.e.,
\begin{equation}
q_0^\mathrm{reco} \simeq \Ehad = \sum_i^\textrm{nucleons} T_i + \sum_j^\textrm{mesons} E_j .
\label{eq:Ehad}
\end{equation}
We include a 30\% smearing on $\Ehad$, which roughly accounts for NOvA's detector response.
Lastly, NOvA tune uses the reconstructed three-momentum transfer, $|\vec{q}\,^\mathrm{reco}|^2=2(\Ehad+E_\mu)(E_\mu+p_\mu \cos\theta_\mu)-m_\mu^2+E_\mathrm{had}^2$, where $E_\mu$, $p_\mu$, $\theta_\mu$ are the energy, momentum, and the angle with respect to the beam of the muon, for $\nu_\mu$ charged-current events. 
After all the modifications to \texttt{GENIE} described above, in the NOvA ND tune, one then compares the simulated event distribution to the measured one in the 2D plane of reconstructed variables
$(|\vec{q}\,^\mathrm{reco}|, q_0^{\rm reco})$ with 20 bins in $|\vec{q}\,^\mathrm{reco}|$ and 16 bins in $q_0^{\rm reco}$. 
With the assumption that all the discrepancies between the simulated and measured distributions are due to mis-modeling of the meson-exchange current (MEC), one multiplies the MEC event rate predicted by \texttt{GENIE} in every bin by a weight, i.e., ``tune,'' such that the tuned prediction matches the measurement. 
The total number of free parameters in this tuning is 200. See Fig.~\ref{Fig:flowchart} for a schematic illustration.

\section{Sterile neutrinos}
\label{sec:steriles}

To estimate the impact of the tuning procedure on BSM searches, we choose to study two illustrative models: eV-scale sterile neutrinos~\cite{Kainulainen:1990bn} and light neutrinophilic scalars~\cite{Kelly:2019wow}.
The reason for choosing these two models is their different experimental signatures.
In this section, we start with sterile neutrinos with masses around the eV scale, for which oscillations of active neutrinos to sterile neutrinos will look like wiggles in the ND neutrino energy spectrum.

\subsection{Model description}

For two decades, the standard three neutrino paradigm has been confronted with experimental anomalies that could indicate oscillations due to a mass splitting much larger than the measured solar and atmospheric ones.
These anomalies are fourfold: LSND~\cite{LSND:2001aii},  
MiniBooNE~\cite{MiniBooNE:2020pnu}, reactor~\cite{Mueller:2011nm, Mention:2011rk, Huber:2011wv}, and gallium~\cite{GALLEX:1997lja, SAGE:1998fvr, Abdurashitov:2005tb, Kaether:2010ag, Giunti:2010zu, Barinov:2021asz} anomalies.
All these anomalies seem consistent with oscillations at a baseline distance much shorter than expected, pointing to a new mass splitting $\Delta m^2_{41}\sim\mathcal{O}(1~{\rm eV}^2)$.
Nevertheless, direct~\cite{ALEPH:1993pqw, OPAL:1994kgw, L3:1998uub} and indirect~\cite{DELPHI:2000wje} measurements of the $Z$ invisible width at LEP would imply that this state does not interact via weak interactions and thus has no standard model gauge quantum number whatsoever: a \emph{sterile neutrino}.

Short baseline oscillations may be approximated by
\begin{equation}
  P(\nu_\alpha\to\nu_\beta)\simeq \delta_{\alpha \beta} - 4|U_{\alpha 4}|^2(\delta_{\alpha \beta}-|U_{\beta 4}|^2)\sin^2\left(\frac{\Delta m^2_{41} L}{4E_\nu}\right),
\end{equation}
where $U_{\alpha i}$ denotes the extended 4$\times$4 PMNS matrix, $\Delta m^2_{41}\equiv m_4^2-m_1^2$, and $E_\nu$ is the neutrino energy.
Explaining the LSND and MiniBooNE anomalies requires $\nu_\mu\to\nu_e$ transitions, and thus a nonzero $U_{e4}$ and $U_{\mu4}$.
This, in turn, implies short baseline $\nu_\mu$ and $\nu_e$ disappearance.
This relation between appearance and disappearance channels leads to a strong tension between data sets under the interpretation of sterile neutrino oscillations~\cite{Dentler:2018sju, Moulai:2019gpi, Giunti:2019aiy}. 
This tension is mainly driven by the following experiments: MINOS/MINOS+~\cite{MINOS:2020iqj} and IceCube\cite{IceCube:2016rnb, IceCube:2020tka} for $\nu_\mu$ disappearance; Daya Bay~\cite{MINOS:2020iqj}, as well as solar~\cite{SNO:2008gqy, Super-Kamiokande:2016yck, BOREXINO:2018ohr}, short baseline~\cite{RENO:2020hva, DANSS:2018fnn}, and radioactive source experiments~\cite{Barinov:2021asz} for $\nu_e$ disappearance; and LSND and MiniBooNE for $\nu_\mu\to\nu_e$ appearance. 

\begin{figure}[t] 
	\centering
	\includegraphics[width=0.45\textwidth]{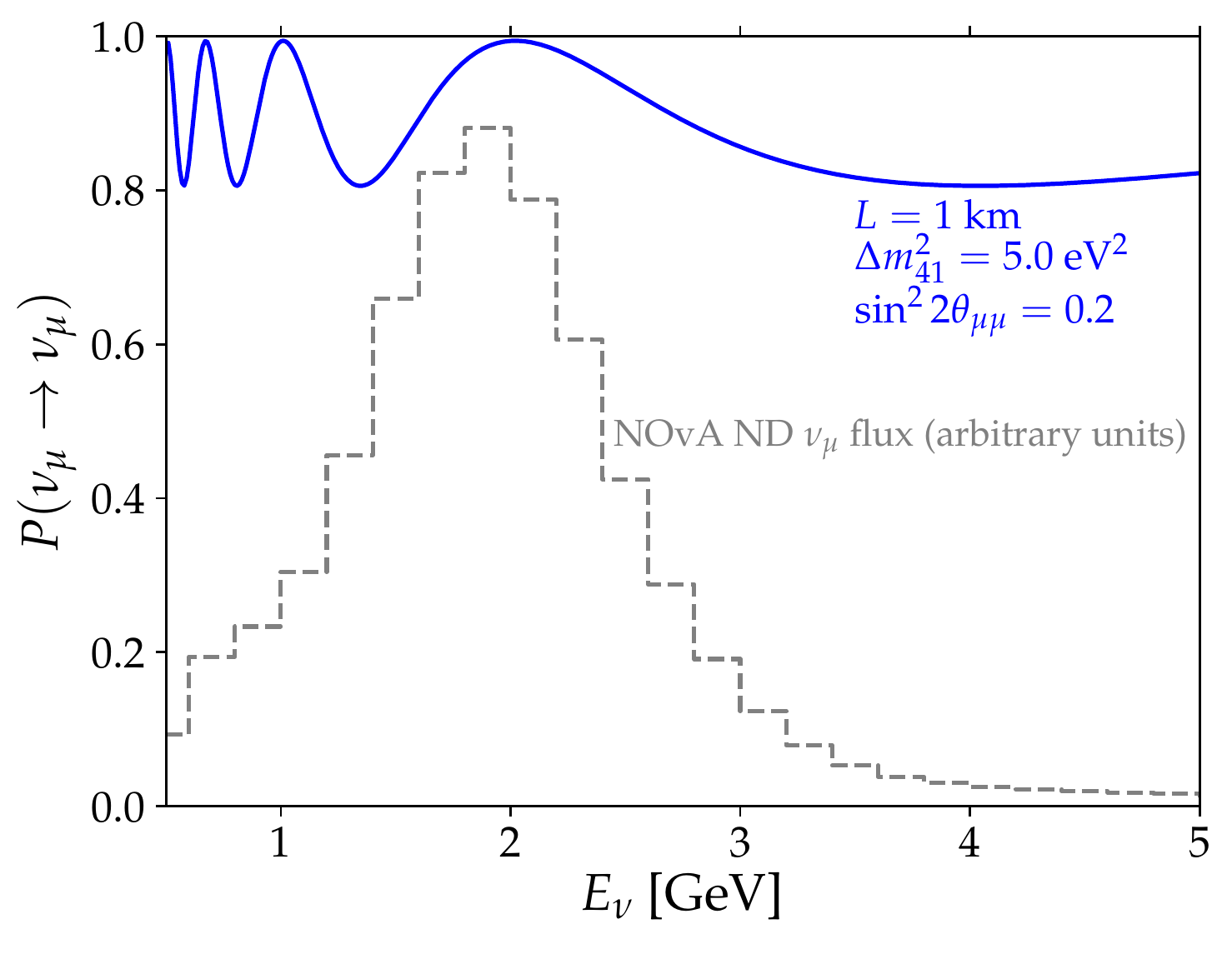}
	\caption{Oscillation probability in the $\nu_\mu$ disappearance channel for illustrative sterile neutrino oscillation parameters as a function of the true neutrino energy $E_\nu$ (blue line). We also show the shape of the NOvA $\nu_\mu$ flux in neutrino mode for reference (gray histogram)~\cite{NOvAflux}.
	\label{Fig:probability}
	}
\end{figure}

Regardless of this tension, testing short baseline oscillations with disappearance channels has become increasingly important in the last decade.
Here, we are interested in the $\nu_\mu$ disappearance channel at NOvA ND for which
\begin{equation}
  P(\nu_\mu\to\nu_\mu)\simeq 1-\sin^2 2\theta_{\mu\mu} \sin^2\left(\frac{\Delta m^2_{41} L}{4E_\nu}\right),
\end{equation}
where we have defined the effective $\nu_\mu$ disappearance angle $\sin^22\theta_{\mu\mu}\equiv 4|U_{\mu4}|^2(1-|U_{\mu4}|^2)$.
Oscillations in NOvA ND, located about 1~km from the beam target, are depicted for illustrative oscillation parameters in Fig.~\ref{Fig:probability}.
Clearly, a sterile neutrino could induce an oscillatory pattern in NOvA's neutrino spectrum at the ND.
The question we want to ask is how much of this oscillatory pattern will be affected by NOvA's tuning procedure, and what the corresponding impact is on NOvA's sensitivity to sterile neutrinos.

\subsection{Analysis}
\label{Sec:ResultsSterile}

Let us start with the general approach used to perform  analyses in this work. 
To mimic a sterile neutrino search in NOvA, we first use an event generator $\mathcal{G}$ to generate mock data for both ND and FD that contain a sterile neutrino signal. We then use either the same or a different event generator $\mathcal{G}'$ to tune a SM event set to the mock data in the ND. 
The idea behind using a different generator to fit the mock data is to account for possible mis-modeling of neutrino-nucleus scattering.
Lastly, we do a chi-square analysis using the mock data and the tuned generator $\mathcal{G}'$ to see if we can recover the sterile signals and to examine how the sensitivity changes because of the tuning.

More concretely, we use the \texttt{GENIE} v2.12.2 event generator to produce mock data. 
We produce a set of $\nu_\mu$-carbon charged current events and reweight them according to the NOvA flux and the oscillation probability $P(\nu_{\mu} \to \nu_{\mu})$ at the ND. 
We account for the varying decay lengths of the pions in the decay pipe. 
Following Ref.~\cite{NOvAdecay}, we take a random oscillation length $L$ from an exponential distribution that has a maximum at $L=1$ km and extends to $L=0.3$ km.
The reconstructed neutrino energy is obtained following the procedure discussed in Sec.~\ref{sec:Methods}: $E_\nu^{\rm reco}=E_\mu+q_0^{\rm reco}$, where $q_0^{\rm reco}$ is the hadronic energy $E_{\rm had}$ smeared by 30\%.

To fit the data, we study several cases. 
We use either \texttt{GENIE} v2.12.2 or \texttt{NuWro} v19.02.2~\footnote{For the MEC component, \texttt{GENIE} adopts the so-called ``Empirical MEC'' or ``Dytman'' model~\cite{Katori:2013eoa}, while \texttt{NuWro} uses the ``Nieves'' model~\cite{Nieves:2011pp}.} to generate the SM spectra, reweighting the events as above.
We fit mock data both with and without performing the tuning procedure to understand the extent of the impact of the tuning, especially its interplay with cross section mis-modeling, on the experimental sensitivity.

For the tuning, we follow NOvA's procedure as discussed in Sec.~\ref{sec:Methods}; however, because we are working with two sets of simulated events, we perform the tune entirely in the true $(|\vec{q}|,q_0)$ plane.
When considering sterile neutrinos in the fit generator $\mathcal{G}'$, one could either first reweight according to the oscillation probability and then tune, or the other way around.
Performing the analysis by first reweighting ensures us that when we use the same generator for both mock data and fit model, and we compare the same parameters for sterile neutrinos, the tuning does not make any changes to the generated events because, by design, the mock data and the fit generator predict the same ND spectra.

\begin{figure}[t!]
	\centering
	\includegraphics[width=0.48\textwidth]{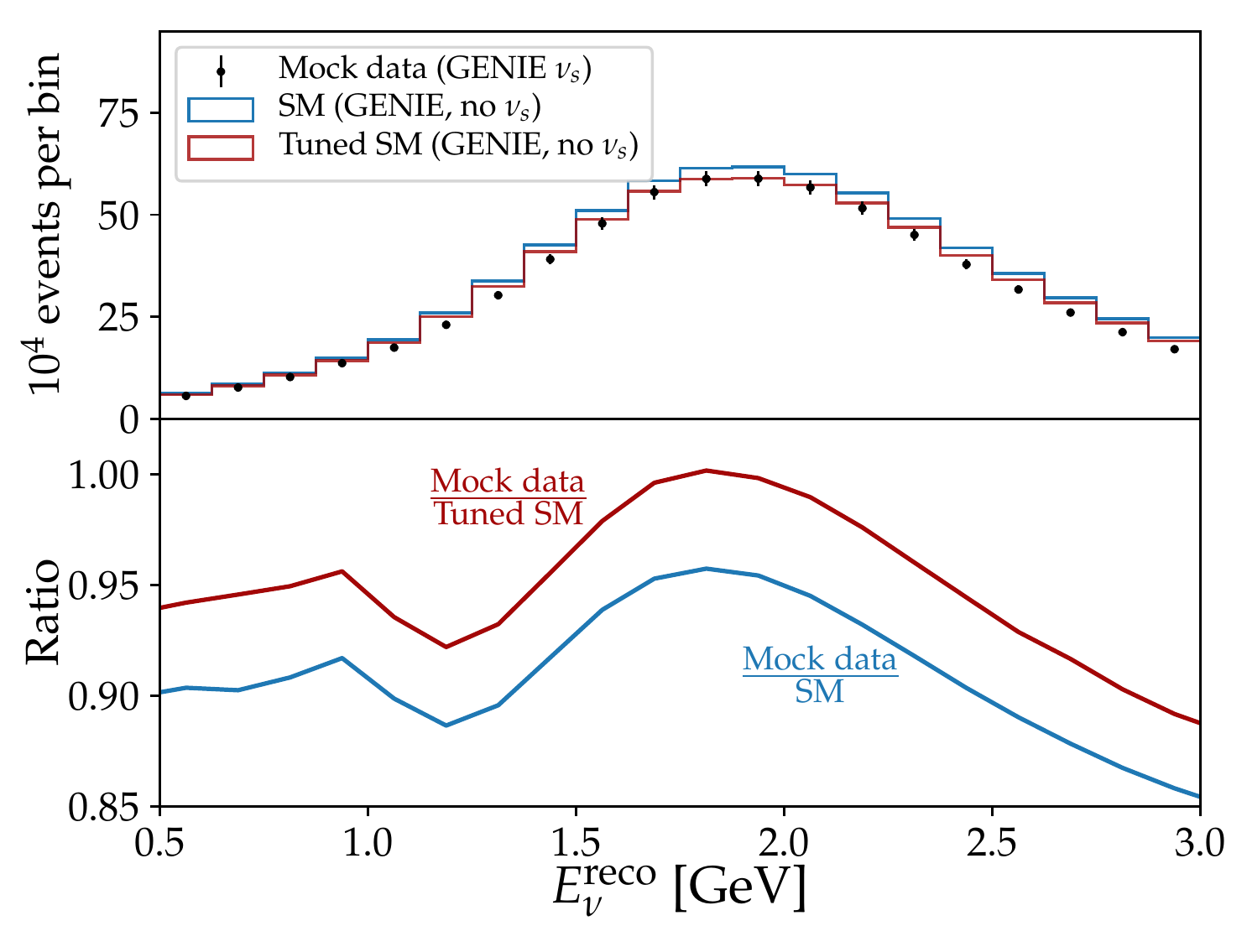}
	\includegraphics[width=0.48\textwidth]{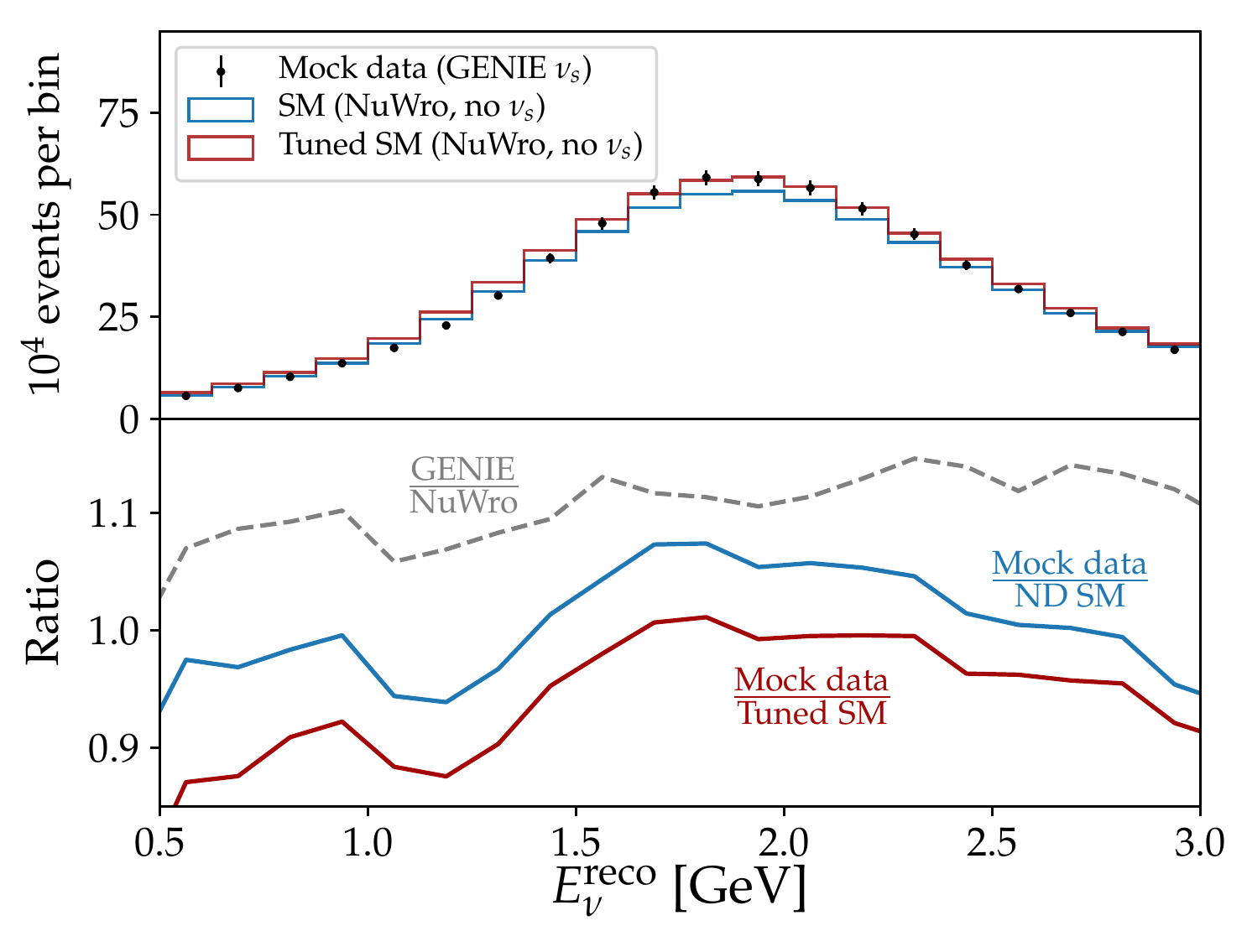}
	\caption{ND distributions of $E_{\nu}^{\mathrm{reco}}$ for sterile data and SM expectation for $\Delta m_{41}^2=5$ eV$^2$ and $\sin^2(2\theta_{\mu\mu})=0.2$. In the left panel, the data and model are both generated using \texttt{GENIE}; in the right panel, the model is generated using \texttt{NuWro}. The top panel shows a histogram of events for the SM (blue), mock data (black points), and tuned SM (red) at the ND, while the lower panel shows the ratio ND data / ND SM for both the tuned (red) and untuned (blue) SM expectation.
	\label{Fig:sterile_dNdE}
	}
\end{figure}

To appreciate the impact of the tuning, we present Fig.~\ref{Fig:sterile_dNdE}.
In the left panel, we compare the mock data (black points) against the fit spectra using \texttt{GENIE} without (blue) and with (red) tuning.
Mock data was generated with a sterile neutrino signal, assuming $\Delta m^2_{41}=5$~eV$^2$ and $\sin^2 2\theta_{\mu\mu}=0.2$, while the fit spectra have no sterile signal.
The top panel shows the spectra and the bottom panel shows the ratios of the data to fit spectra. The blue lines reflect the situation that one would intuitively expect in a sterile neutrino analysis. 
In the top panel,  we see that the data rate is lower than the SM rate because of the existence of a sterile neutrino;  in the bottom panel, the blue line shows the characteristic oscillatory feature of a sterile neutrino signal. 
The wiggles look slightly different from Fig.~\ref{Fig:probability} due to energy and baseline smearing. The red lines, meanwhile, illustrate the effect of tuning. The top panel shows that the tuning does not ``fix'' all the discrepancies between the fit generator prediction and the data. This might be surprising because the tuning has many degrees of freedom; however, the tuning is performed in  the $(\vec{q}, q_0)$ plane, and  these observables do not correlate with $E_\nu$ in a straightforward way. In fact, the red line in the bottom panel shows that if our modeling of the SM neutrino-carbon interactions is perfect (when we use \texttt{GENIE} to model both the data and the fit spectra), the tuning almost perfectly preserves the sterile neutrino feature. 
For more detailed discussions, see Appendix~\ref{subsec:steriles}.

The right panel of Fig.~\ref{Fig:sterile_dNdE} shows the same mock data but with the fit generator being $\texttt{NuWro}$. This is a more realistic situation, where our modeling of neutrino-carbon interactions, i.e., by the fit generator $\texttt{NuWro}$, is different from the true interactions, i.e., by the mock data generator $\texttt{GENIE}$. To mimic the situation in NOvA ND, in which the pre-tune model (\texttt{NuWro}) predicts fewer overall events than the data (\texttt{GENIE}), we rescale the model weights relative to the data by an overall factor of 0.9. For reference, we plot in the bottom panel the ratio between SM $\texttt{GENIE}$ and SM $\texttt{NuWro}$ spectra (grey dashed line). We can see from the red and blue lines in the bottom panel that some oscillatory features are still present in the ratio between mock data and fits, but less prominent than in the left panel. This can be partially attributed to the intrinsic differences between baseline generators, without the presence of any new physics, which can be seen in the grey dashed line. We can already see that this may cause a bias in the experimental sensitivity to sterile neutrinos.

To estimate NOvA's sensitivity, we build a covariance matrix, accounting for the following systematics: 20\% overall normalization, 4\% normalization of ND relative to FD, 2\% correlated near-to-far spectral uncertainty (bin-by-bin), and fully uncorrelated 2\% bin-to-bin uncertainty on both near and far detectors. 
Our covariance matrix is meant to capture the general features of NOvA's sensitivity, and not to reproduce precisely the experimental results.
The sensitivity is estimated by a chi-square function $\chi^2=(D-F)\cdot C^{-1}\cdot (D-F)$, where $D$ and $F$ are the data and fit spectra, and $C$ is the covariance matrix including statistical uncertainties.

\begin{figure}[t!]
	\centering
	\includegraphics[width=0.45\textwidth]{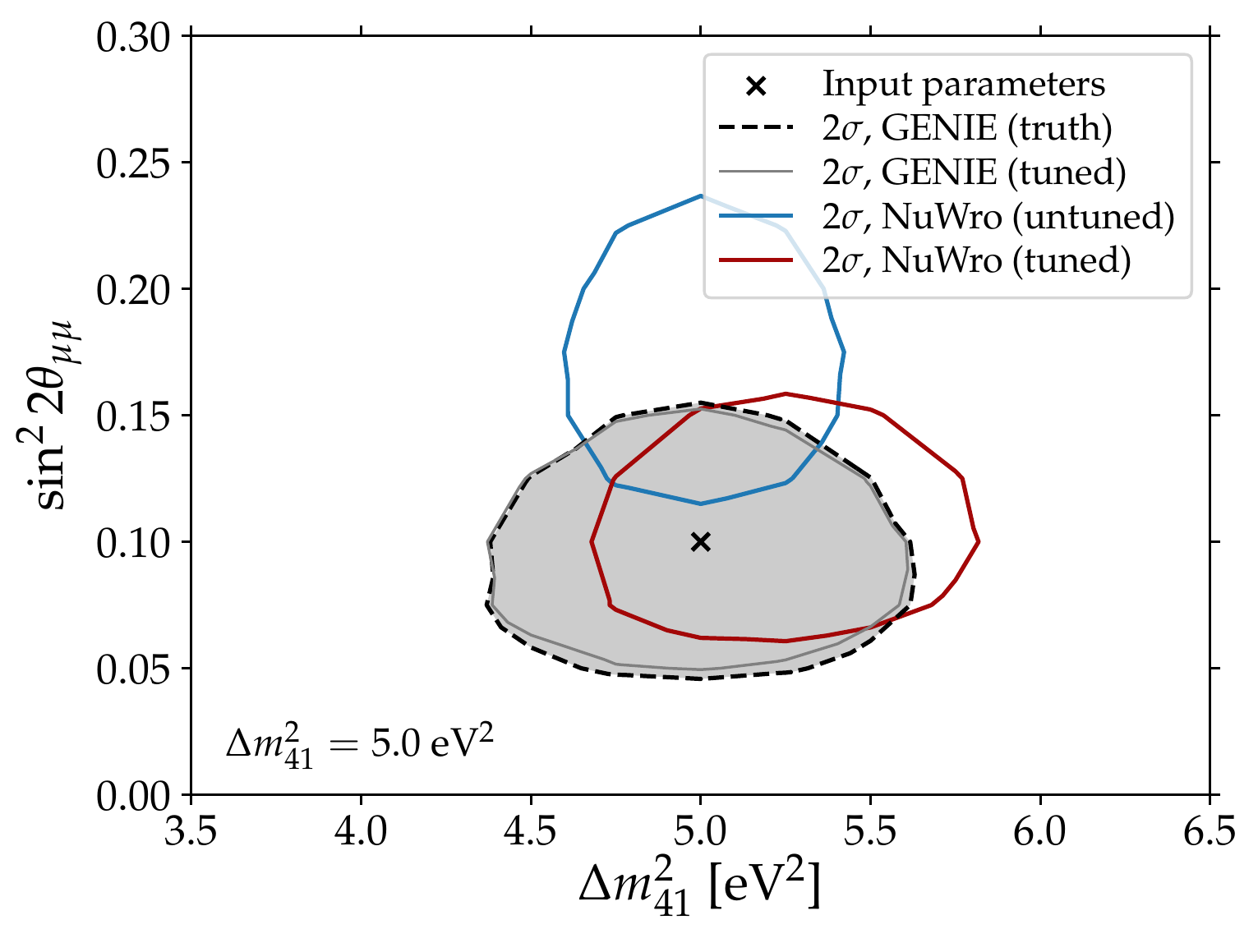}
	\includegraphics[width=0.45\textwidth]{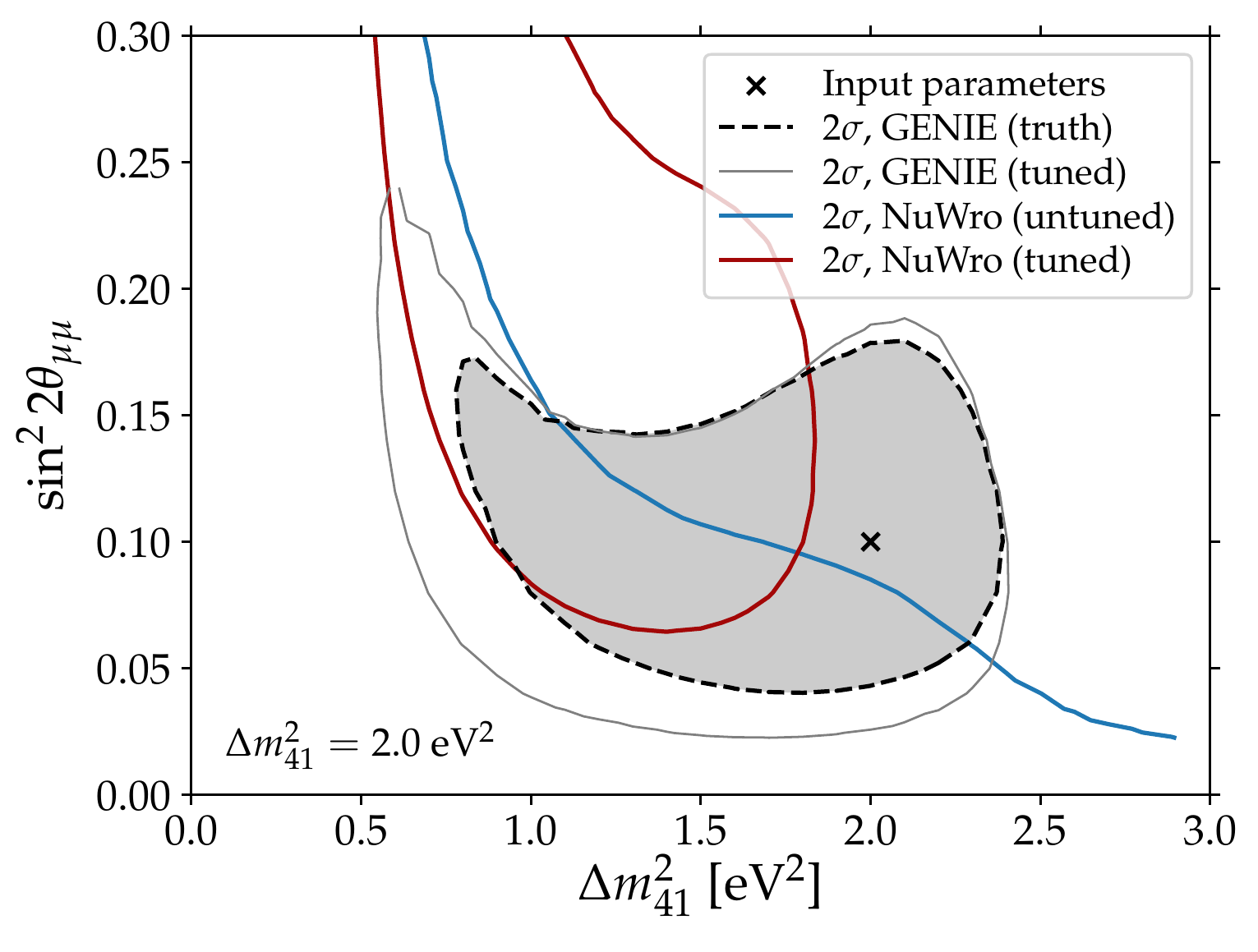}\\
	\caption{Sensitivities to sterile neutrinos for different choices of event generators and tuning, assuming sterile neutrino parameters $\sin^2 2\theta_{\mu\mu}=0.1$ and $\Delta m^2_{41}=5$~eV$^2$ (left panel) and $\Delta m^2_{41}=2$~eV$^2$ (right panel).
	\label{Fig:sterile_chisq}}
\end{figure}

We perform a sensitivity analysis for the tuned and untuned case after adding a sterile neutrino signal with $\sin^2 2\theta_{\mu\mu} = 0.1$ and two choices of mass splittings, $\Delta m_{41}^2 =5.0$ eV$^2$ and 2.0 eV$^2$, as shown in the left and right panels of Fig.~\ref{Fig:sterile_chisq}. 
Again, we generate the mock data with \texttt{GENIE} and estimate the sensitivities for four cases: \texttt{GENIE} without tuning (shaded region), \texttt{GENIE} with tuning (thin line), \texttt{NuWro} without tuning (blue), and \texttt{NuWro} with tuning (red). 
Let us start with the left panel. 
The grey shaded region indicates an ideal analysis, where our fit generator and the data generator are the same, and there is no tuning. 
In this case, we recover the true input parameter, and the systematic uncertainties determine the region size. 
Note that because the FD has very few events, it contributes very little to the $\chi^2$, and the grey region would look almost the same if we did a ND-only analysis. 
We also show the impact of the tuning procedure if one had the correct model of neutrino-carbon interaction (thin line).
We can see that tuning indeed enlarges the region  slightly, but the changes are small. 
The blue region shows that when we mis-model neutrino-carbon interaction and we do not tune our generator, we would still identify a sterile neutrino signal, but the fit parameters would be biased by more than 2$\sigma$. 
Lastly, the red region shows that if we adopt the ND tuning, we improve the accuracy of the fitted parameters. 
This could seem to indicate that the tuning procedure makes the sensitivity more robust to mis-modeling.

The right panel of Fig.~\ref{Fig:sterile_chisq} tells a different story. With just a different $\Delta m_{41}^2$, we see that neither the untuned nor the tuned regions include the real input values at 2$\sigma$. The untuned region is consistent with no sterile neutrinos, and the tuned region prefers large mixing angles. 
This illustrates that cross section mis-modeling can significantly bias the experimental sensitivities and that the tuning procedure does not fix this issue.

\section{Light neutrinophilic scalars}
\label{sec:Mononu}

The second scenario we analyze is the light neutrinophilic scalar model proposed in Refs.~\cite{Berryman:2018ogk, Kelly:2019wow}.
Its experimental signature consists of an excess of missing transverse momentum.
This is a good example of a class of models in which the experimental signature depends on certain aspects of the interaction kinematics on an event-by-event basis. The correlations between these kinematic variables and the tuning variables $q_0$ and $|\vec{q}|$ can be nontrivial. By studying a representative signature, we hope to uncover common lessons for these kinematic searches.

\subsection{Model description}

Light neutrinophilic scalars could mediate interactions between neutrinos and dark matter candidates.
These scalars would couple to neutrinos via the effective operator,
\begin{equation}
\label{eq:mononu_operator}
  \mathcal{O} = \frac{(L_\alpha H)(L_\beta H)}{\Lambda_{\alpha \beta}^2}\phi+{\rm h.c.}\to \frac12 \lambda_{\alpha \beta}\nu_\alpha \nu_\beta\phi+{\rm h.c.},
\end{equation}
where $\phi$ is the neutrinophilic scalar itself; $H$ and $L_\alpha$ are the Higgs and lepton doublets; $\alpha,\beta=e,\mu,\tau$; $\Lambda_{\alpha \beta}$ is the scale of the dimension-6 operator; $\lambda_{\alpha \beta} = v^2/\Lambda_{\alpha \beta}^2$ is an effective coupling between the $\phi$ and neutrinos after electroweak symmetry breaking; and lastly, $v=246$~GeV is the Higgs vacuum expectation value.
A sum over flavor indices is implicit.
In addition to this neutrino interaction term, $\phi$ could also couple to dark matter, so it serves as a portal between neutrinos and the dark sector.

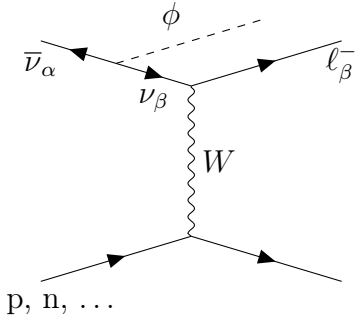
\begin{figure}[t] 
\centering
\begin{tikzpicture}
\begin{feynman}
\vertex (f1);
\vertex[right=2cm of f1] (x1);
\vertex[below=0.6cm of x1] (f2);
\vertex[right=4cm of f1] (f3);
\vertex[right=1cm of f1] (xs1);
\vertex[below=0.3cm of xs1] (s1);
\vertex[right=2cm of s1] (xs2);
\vertex[above=0.6cm of xs2] (s2);
\vertex[below=3.2cm of f1] (f4);
\vertex[right=2cm of f4] (x2);
\vertex[above=0.6cm of x2] (f5);
\vertex[right=4cm of f4] (f6);
\diagram*{
(s1)--[fermion] (f1),
(s1)--[fermion] (f2) --[fermion] (f3),
(f4) --[fermion] (f5)--[fermion] (f6), 
(f2)--[photon, edge label=\(W\)] (f5),
(s1)--[dashed, edge label=\(\phi\)] (s2),
};
\node at (0,-0.3) {$\overline\nu_\alpha$};
\node at (1.5,-0.8) {$\nu_\beta$};
\node at (4,-0.3) {$\ell_\beta^-$};
\node at (0.3,-3.5) {p, n, \dots};
\end{feynman}
\end{tikzpicture}
	\caption{Feynman diagram relevant to the mono-neutrino signal. 
	Note that the hadronic current can go beyond nucleon level (e.g. two-body currents, deep inelastic scattering), so we leave it general.
	\label{Fig:feynman_diagram}
	}
\end{figure}

In this scenario, when a neutrino scatters via charged current interactions, it may radiate a $\phi$, which would either leave the detector or decay invisibly to neutrinos or dark matter, thus constituting missing momentum.
From Eq.~\eqref{eq:mononu_operator}, we note that $\phi$ carries lepton number; thus, when emitted, $\phi$ changes a neutrino to an antineutrino and vice versa.
We depict the Feynman diagram inducing this transition in Fig.~\ref{Fig:feynman_diagram}.
Note that we have not been very explicit on the hadronic current, as the neutrino may scatter on nucleons, quarks, and even two-body currents.
This signature has been named the \emph{mono-neutrino} signal, in tandem with collider mono-$X$ searches~\cite{Abercrombie:2015wmb}.

The typical experimental signature of this scenario will be an excess of missing transverse momentum, $\slashed{p}_T$, compared to usual neutrino scattering events. The missing transverse momentum is defined as the magnitude of the sum of the transverse momenta of the visible particles,
\begin{equation}
\slashed{p}_T = \left| \sum_{\mathrm{i,vis}} (\vec{p}_T)_i \right| .
\end{equation}
While the $\phi$ emission will necessarily lead to a missing momentum, standard neutrino-nucleus interactions could also display relatively large missing transverse momenta.
There are three main reasons for that.
If a neutrino interacts with a nucleon inside a nucleus, the struck nucleon is not at rest due to Fermi motion. 
The Fermi momentum is typically of the order of 250~MeV or so, leading to an unavoidable spectrum of $\slashed{p}_T$.
Moreover, neutrinos scatter with particles inside a dense nuclear medium.
As particles propagate throughout the nucleus, secondary scatterings may occur, referred to as final state interactions.
This effect is fairly common and may lead to large $\slashed{p}_T$. Lastly, outgoing hadrons, particularly neutrons, could escape detection and thus lead to $\slashed{p}_T$.

\subsection{Analysis}
\label{Sec:ResultsMononu}

To simulate a set of mono-neutrino events, we use MadGraph5~\cite{Alwall:2011uj} to generate $\overline\nu_{\mu} + n \to \mu^- + \phi + p$ events, treating the proton and neutron as elementary particles. 
Although this neglects Fermi motion and final state interactions, the emission of a $\phi$ dominates the missing transverse momentum distribution.
We choose to work with an antineutrino beam because the final-state protons are visible and can be used to reduce SM backgrounds, as we will discuss later.
We generate events at fixed neutrino energies, then weight the events by the convolution of the mono-neutrino cross section and the NOvA antineutrino flux at the given energy for each event.
In this analysis, we use \texttt{NuWro} to simulate SM interactions, which accounts for Fermi motion and final state interactions. 
Here we treat neutrons as missing energy and include a 30\% smearing on hadronic momenta.
Mock data is obtained by combining the signal and background events, producing a simulated expected missing momentum spectrum at the ND for this model.
To account for mis-modeling of neutrino-nucleus interactions, we use \texttt{GENIE} to simulate our SM fit background.

\begin{figure}
	\centering
	\includegraphics[width=0.45\textwidth]{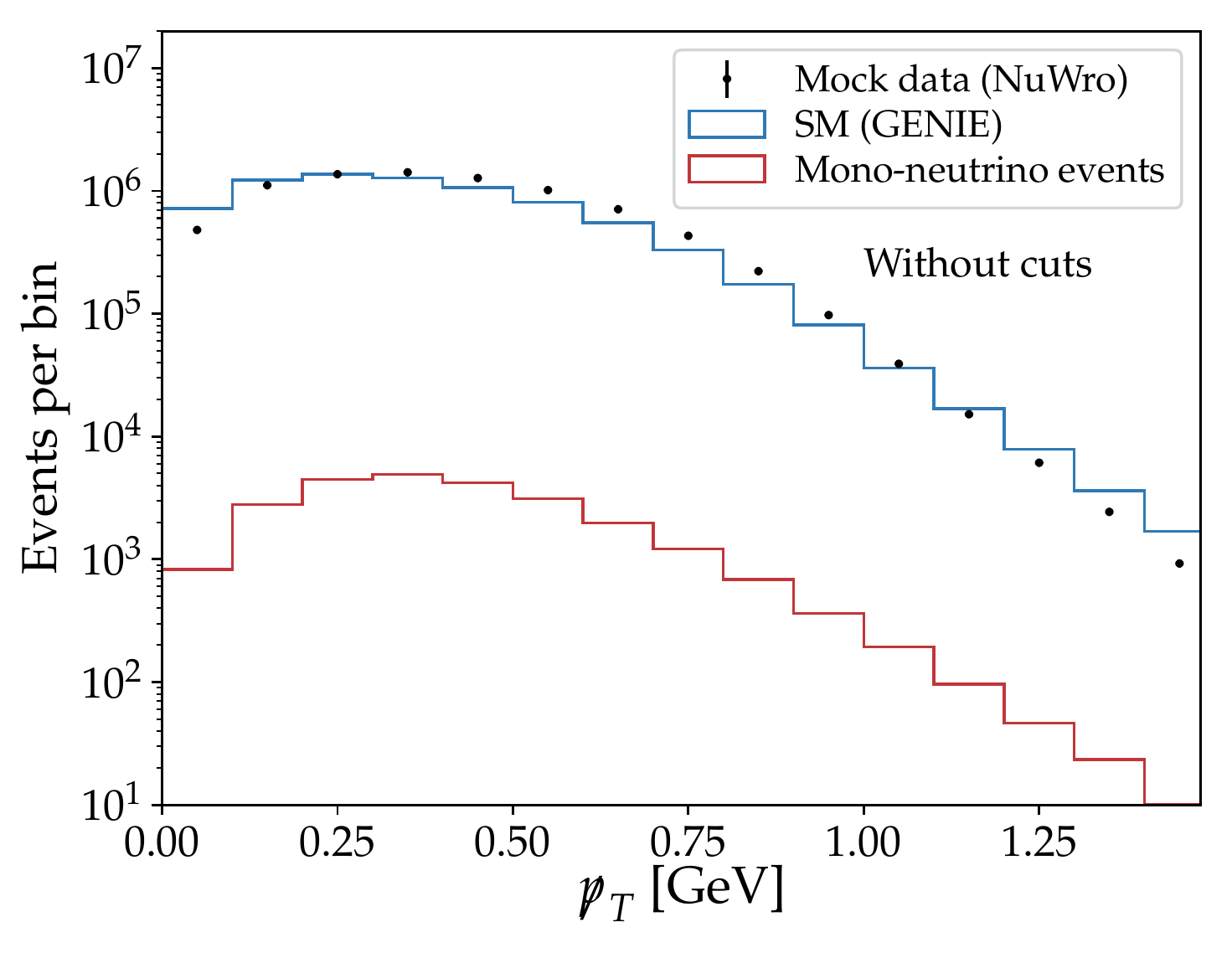}
	\includegraphics[width=0.45\textwidth]{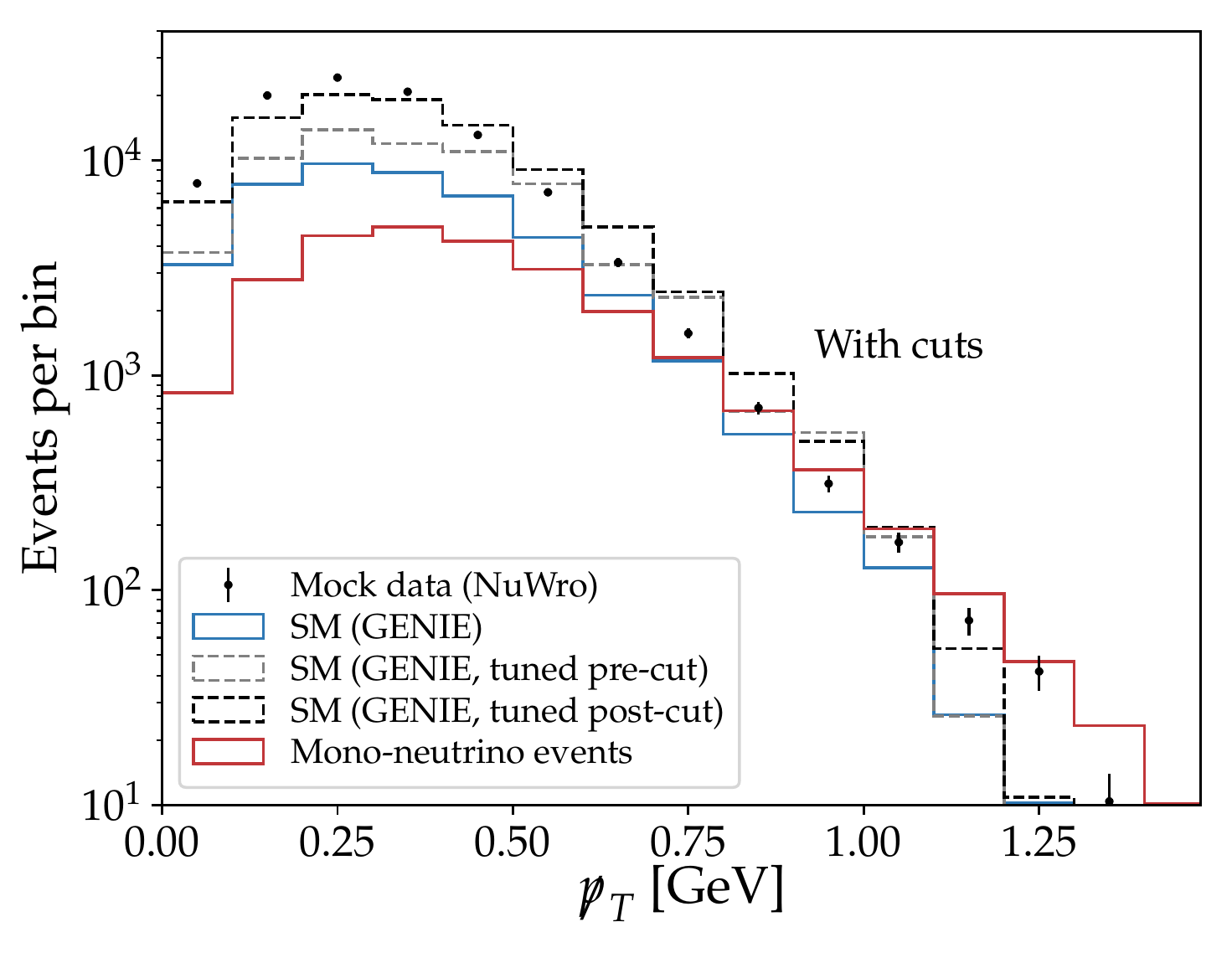}
	\caption{\label{fig:mononu}
		Missing $p_T$ distributions before (left) and after (right) cuts. 
		Mock SM data (black points) was generated with \texttt{NuWro}, while \texttt{GENIE} was used as fit model (blue). 
		The mono-neutrino signal is given in red.
		In the right panel we also show the predicted $/\!\!\!p_T$ spectra for tuned fit model when the tuning is performed before cuts (gray) or after cuts (black).}
\end{figure}

In the left panel of Fig.~\ref{fig:mononu}, we show the missing momentum spectra for the mock data background (black points), the fit model (blue), and the mono-neutrino signal (red) for coupling $\lambda_{\mu\mu}=0.5$ and scalar mass $m_\phi=0.5$~GeV.
As we can see, the mono-neutrino signal is more than two orders of magnitude below the background.
This is partially due to the multiplicity of the final states of the signal. 
The extra phase space factor of three- compared to two-body final states leads to a suppression of about two orders of magnitude on the cross section.
This low signal rate forces us to impose experimental cuts to improve the signal-to-background ratio.

In antineutrino mode, the signal events are given by $\overline\nu_\mu + n\rightarrow \mu^- +\phi + p$. The visible final states in the detector are the muon and the proton. In contrast, we can think of the background events naively as given by $\overline\nu_\mu +p\rightarrow \mu^+ +n$. 
The presence of a neutron could lead to large $\slashed{p}_T$, which would mimic the signal.
While neutrons are hard to reconstruct, those above 100 MeV are likely to have hard interactions with nuclear matter, leading to hadronic activity that allows us to identify its presence.
Given this, we cut events with neutrons above 100 MeV.~\footnote{This is consistent with setting all the neutron energy as missing energy. We assume that, even though NOvA can identify the presence of a neutron above 100 MeV, they cannot reconstruct its energy, and therefore its energy still goes to missing energy.}
Moreover, we only select events with exactly one visible proton, that is, protons with kinetic energy above 100 MeV, and no pions at all.
We include a 30\% smearing on hadronic momenta.

The result of these cuts can be seen in the right panel of Fig.~\ref{fig:mononu}.
While the signal-to-background ratio has improved considerably, a new issue arises: the background-only mock data is significantly different from the fit model (blue).
Under further consideration, this should not be surprising because the more exclusive we look at cross section predictions, the larger the expected theoretical uncertainties and discrepancies. 
To understand the extent to which this large discrepancy could be addressed by ND tuning, we perform the tuning before (gray) and after (black) experimental cuts.
We can clearly see that neither options can reproduce the mock data spectrum.
But, more importantly, the tuned spectra depend strongly on the cuts used, so even if a tune is performed with one event sample and measured with another, this dependence may introduce biases in the experimental analyses.
On top of that, the differences between tuned model and data are larger than the signal rate itself. 
For more details on discrepancy between generators and the impact of the tuning to mono-neutrino signal, see Appendix~\ref{subsec:mononu}.

This large discrepancy between mock data and tuned fit model already shows us that the mis-modeling of neutrino-nucleus interactions may render an experiment completely unable to probe certain BSM physics scenarios.
Regardless, we can perform a statistical analysis of the experimental sensitivity to appreciate quantitatively how relevant cross section modeling is to this BSM search.
We perform the analysis by employing a similar covariance matrix to that built in the previous section, but excluding the FD data, and with the uncertainties now applicable to bins of $\slashed{p}_T$.
For systematics we take an overall normalization uncertainty of 20\% and a bin-to-bin uncorrelated uncertainty of 5\%; our chi-square is $\chi^2 = (D-F)\cdot C^{-1} \cdot(D-F)$, where $D$ and $F$ are the data and fit model binned in $\slashed{p}_T$.
While this is not a realistic treatment of experimental sensitivity to missing momentum signals, it suffices for our purposes.

\begin{figure}
	\centering
	\includegraphics[width=0.45\textwidth]{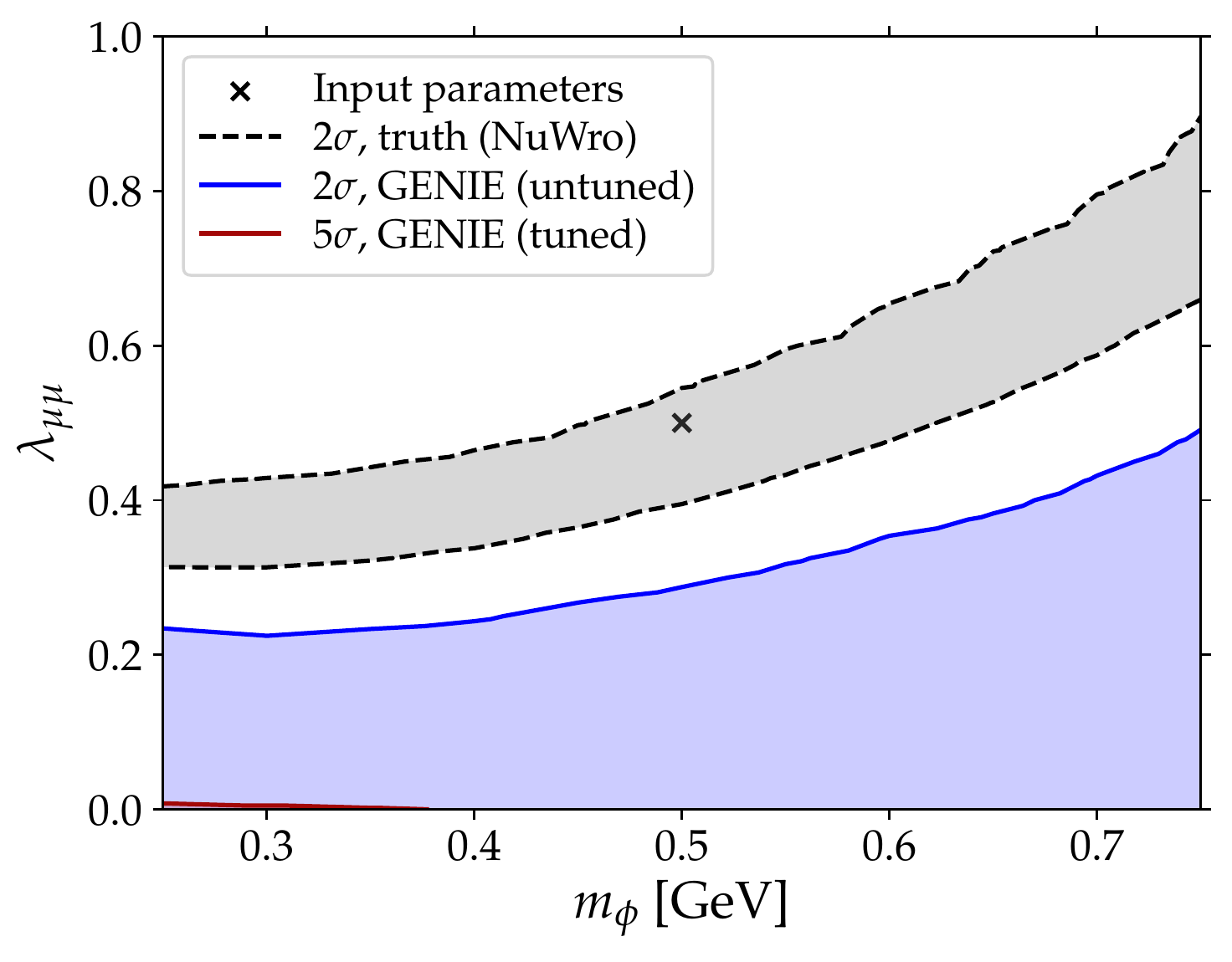}
	\caption{\label{fig:mononu_sensitivity}
		Sensitivity for a mono-neutrino signal with $\lambda_{\mu\mu} = 0.5$ and $m_{\phi}=0.5$ GeV for different choices of event generators and tuning. Because the $2\sigma$ tuned region is not visible on the figure, we show the $5\sigma$ region (in the lower left corner).}
\end{figure}

Figure~\ref{fig:mononu_sensitivity} shows our results, assuming the presence of a mono-neutrino signal with coupling $\lambda_{\mu\mu}=0.5$ and $m_{\phi}=0.5$~GeV. In grey, we present the allowed region when using the same generator for the signal and the fit model, which would correspond to the case of perfect modeling of neutrino-nucleus interactions. We also show the allowed region for data and fit model using different generators, which would correspond to cross section mis-modeling. We see that without tuning (blue), the fit prefers no new physics, as the allowed region includes $\lambda_{\mu\mu}=0$. In fact, the fit excludes the true parameter point at more than 5$\sigma$. With tuning (red), the results actually get worse, as the fit strongly prefers no new physics. We can conclude that the discrepancy between mock data and the tuned fit model dominates the experimental sensitivity.

This shows that while near detectors have the potential to probe new physics, without proper modeling of neutrino-nucleus interactions, we may lose this potential. Tuning is not the solution.

\section{Discussions and conclusions}
\label{sec:Conclusions}

In this paper, we have estimated the impact of the ND neutrino-nucleus cross section tuning procedure on BSM signatures at neutrino experiments.
To perform a realistic study, we follow very closely the NOvA collaboration tuning procedure, and we simulate neutrino scattering events with  state-of-the-art event generators \texttt{GENIE} and \texttt{NuWro}.
We study two illustrative BSM scenarios: eV-scale sterile neutrinos, which would manifest via wiggles in the $\nu_\mu$  survival oscillation probability; and neutrinophilic scalars, which could be emitted by neutrinos in charged current interactions leading to large amounts of missing transverse momentum.

Our results show that the interplay between cross section mis-modeling, ND tuning, and sensitivity to new physics is far from trivial.
Concretely, the wiggles in the ND neutrino energy spectrum induced by eV-scale sterile  neutrino oscillations largely remain after tuning.
That is, the tuning of the neutrino-nucleus cross section cannot mimic sterile neutrino signatures.
Nevertheless, mis-modeling neutrino-nucleus interactions can lead to discrepancies between theory and data that may introduce a bias on the experimental sensitivity. 

The situation for neutrinophilic scalars is quite different.
First, the beyond standard model signature is significantly lower than the standard model background.
This forces us to impose experimental cuts to improve the signal-to-background ratio.
In doing so, we go to regions of parameter space in which the background is suppressed but most of the signal remains; however, in these regions the tuning of the fit model is  inefficient at reproducing the data, leading to large discrepancies between data and theory.
We have seen that for the case of neutrinophilic scalars, this discrepancy dominates the experimental sensitivity, rendering it unrealistic.

In both BSM scenarios studied here, the mis-modeling of neutrino-nucleus interactions can lead to significant biases on the experimental sensitivity, regardless of the tuning procedure. 
One may wonder if this is a widespread issue in BSM searches at neutrino experiments, such as T2K, HK, and DUNE. 
It is reasonable to presume that a tuning based on the correct physics would be able to resolve mis-modeling; however, ad hoc tunings will most likely fail to capture the correct physics of neutrino-nucleus interactions, e.g. by lacking important correlations or by being arbitrarily limited to specific processes.
Without the correct physics, a tuning will only be able to reproduce experimental data \emph{by chance}.
Agreement between an ad hoc tuning and data in one observable does not necessarily imply that any other observable will be well described by the tuning, as can be seen for instance in recent MINERvA analyses~\cite{MINERvA:2020zzv}.
It should therefore not come as a surprise that a tuning that is not correlated with a BSM signature and thus does not wash it out is also not capable of addressing mis-modeling.

One may think that practically, a conservative approach would be to assign a large systematic error that includes the possibility that there is indeed gross mis-modeling of the cross sections. However, this does not change the fact that significant cross section mis-modeling and therefore large systematic errors would render the ND blind to new physics.
Our findings reveal the importance of properly accounting for the tuning procedure, as well as properly estimating cross section uncertainties for BSM searches
This will be even more relevant for the future experiments HK and DUNE due to their high statistics.
We hope that this will motivate both theorists and experimentalists to carefully consider the interplay between ND tuning and BSM searches.


\begin{acknowledgments}
\noindent We would like to thank Andr\'{e}  De Gouv\^{e}a, Andrew Furmanski, Joachim Kopp, Ivan Martinez-Soler, Jonathan Paley, Afroditi Papadopoulou, Bryan Ramson, Noemi Rocco, Alexandre Sousa, and Jeremy Wolcott for discussions, and especially Kevin Kelly for providing the mono-neutrino UFO files. Fermilab is operated by the Fermi Research Alliance, LLC under contract No. DE-AC02-07CH11359 with the United States Department of Energy. This project has received support from the European Union’s Horizon 2020 research and innovation programme under the Marie Skłodowska-Curie grant agreement No 860881-HIDDeN. This research was supported in part by the National Science Foundation under Grant No. NSF PHY-1748958. The work of N.C.\ at Fermilab was supported  by the U.S. Department of Energy, Office of Science, Office of Workforce Development for Teachers and Scientists, Office of Science Graduate Student Research (SCGSR) program, which is administered by the Oak Ridge Institute for Science and Education (ORISE) for the DOE. ORISE is managed by ORAU under contract number DE-SC0014664. The work of N.C.\ at the University of Chicago has been supported by the DOE grant DE-SC0013642.

\end{acknowledgments}

\newpage
\appendix
\setcounter{figure}{0}
\renewcommand{\thefigure}{A\arabic{figure}}
\renewcommand{\theHfigure}{A\arabic{figure}}

\section{Tuning details}
\label{sec:Appendix}

\begin{figure}[t!]
	\centering
	\includegraphics[width=0.45\textwidth]{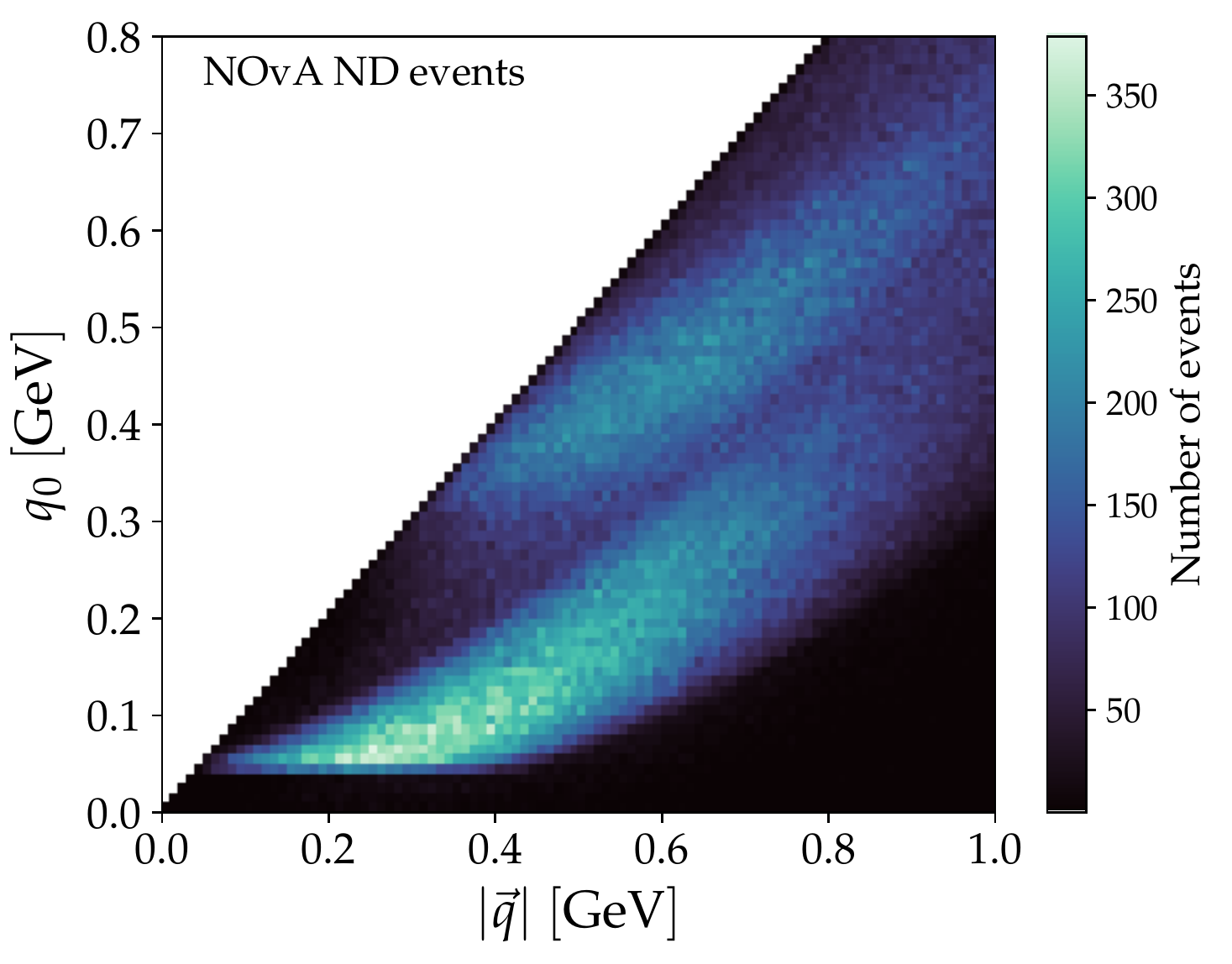}
	\includegraphics[width=0.45\textwidth]{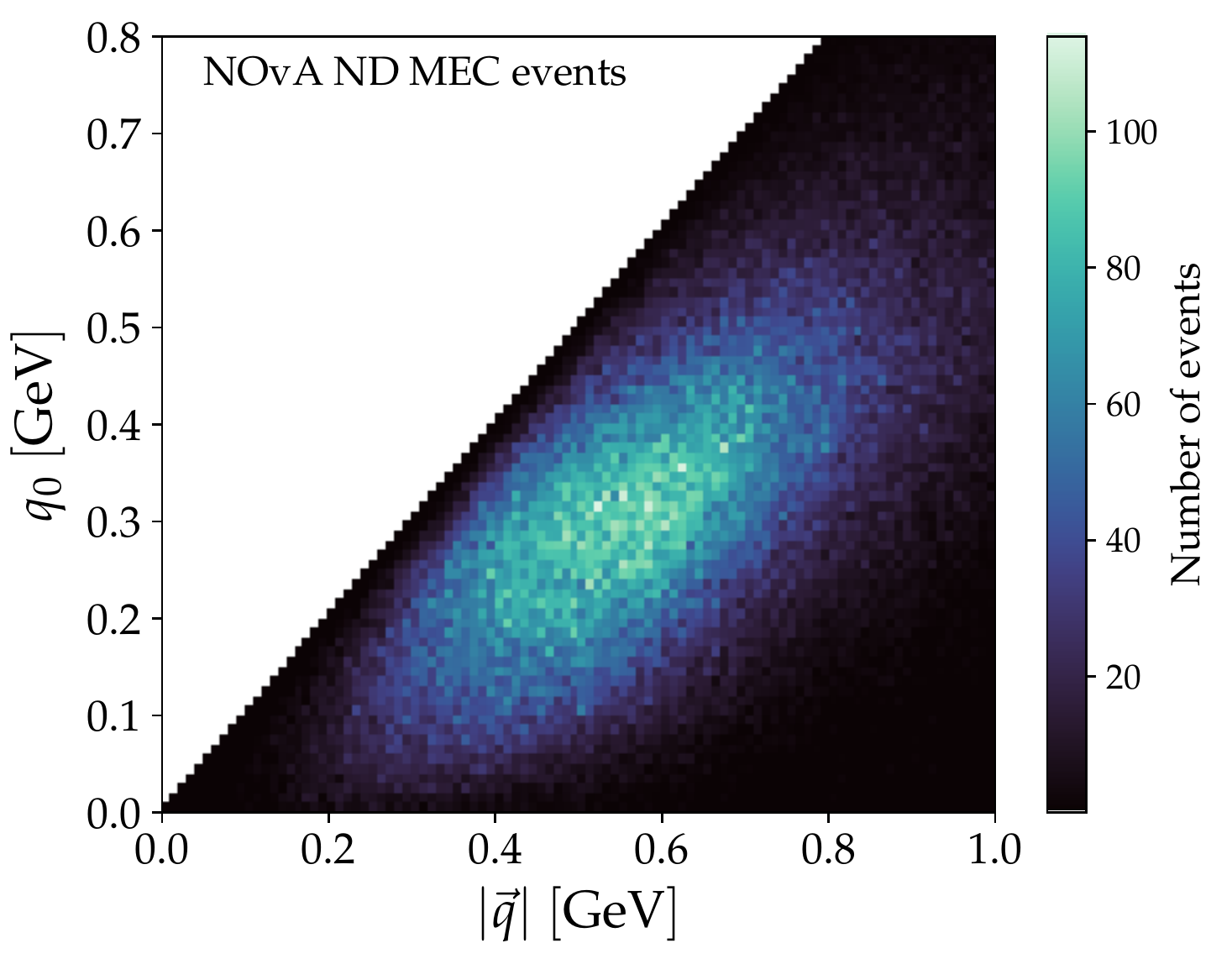}
	\caption{Distributions of all $\nu_\mu$-carbon charged current scattering events (left) and MEC events (right) in NOvA ND in the true $(|\vec{q}\,|,\,q^0)$ plane.
	In the left panel, the lower strip is dominated by quasi-elastic events, while the upper strip comes mainly from resonant and deep inelastic events.
		\label{Fig:SM_pretune}}
\end{figure}

In this appendix, we provide further details on the tuning procedure and its impact on the two observables in which we are interested: reconstructed neutrino energy and missing transverse momentum.
As discussed in Sec.~\ref{sec:Methods}, the NOvA tuning is performed in the $(|\vec{q}\,|,\,q^0)$ plane, only for MEC events. 
While this tuning has many parameters (200 bins), it has two main limitations.
First, $|\vec{q}\,|$ and $q^0$ may not be fully correlated with the relevant experimental observables in an analysis.
Second, since the tuning is applied only to the MEC component of the cross section, it does not have full coverage of the $(|\vec{q}\,|,\,q^0)$ plane.
The latter point can be appreciated in Fig.~\ref{Fig:SM_pretune}, where we show the  distributions of $\nu_\mu$ charged-current events in NOvA ND for all events (left panel) and for MEC events only (right panel), according to the \texttt{GENIE} event generator.
The NOvA ND flux was taken from Ref.~\cite{NOvAflux}. 

In the left panel, the brightest strip is dominated by quasi-elastic scattering events, while the lighter strip above has mostly resonance production and some deep-inelastic scattering events. 
The top triangular area is not kinematically accessible simply because charged current scattering on nuclei proceeds via $t$-channel, and thus $q^2 = (q^0)^2-|\vec{q}\,|^2 < 0$.
Note also that the weights applied in the tuning of the MEC cross section cannot be negative, further limiting the coverage of the tuning procedure.

\subsection{Sterile neutrinos}
\label{subsec:steriles}

\begin{figure}[t!]
	\centering
	\includegraphics[width=0.45\textwidth]{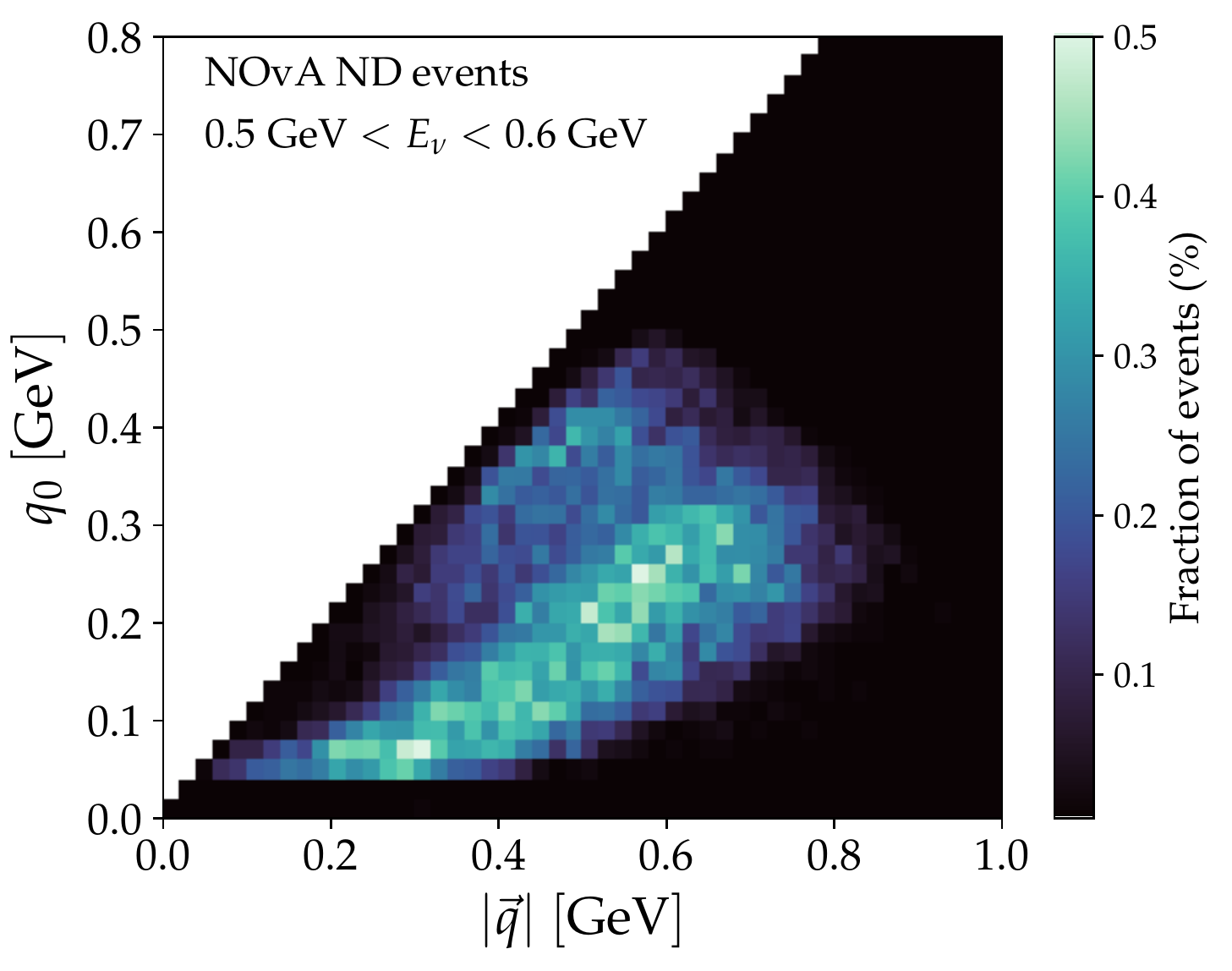}
	\includegraphics[width=0.45\textwidth]{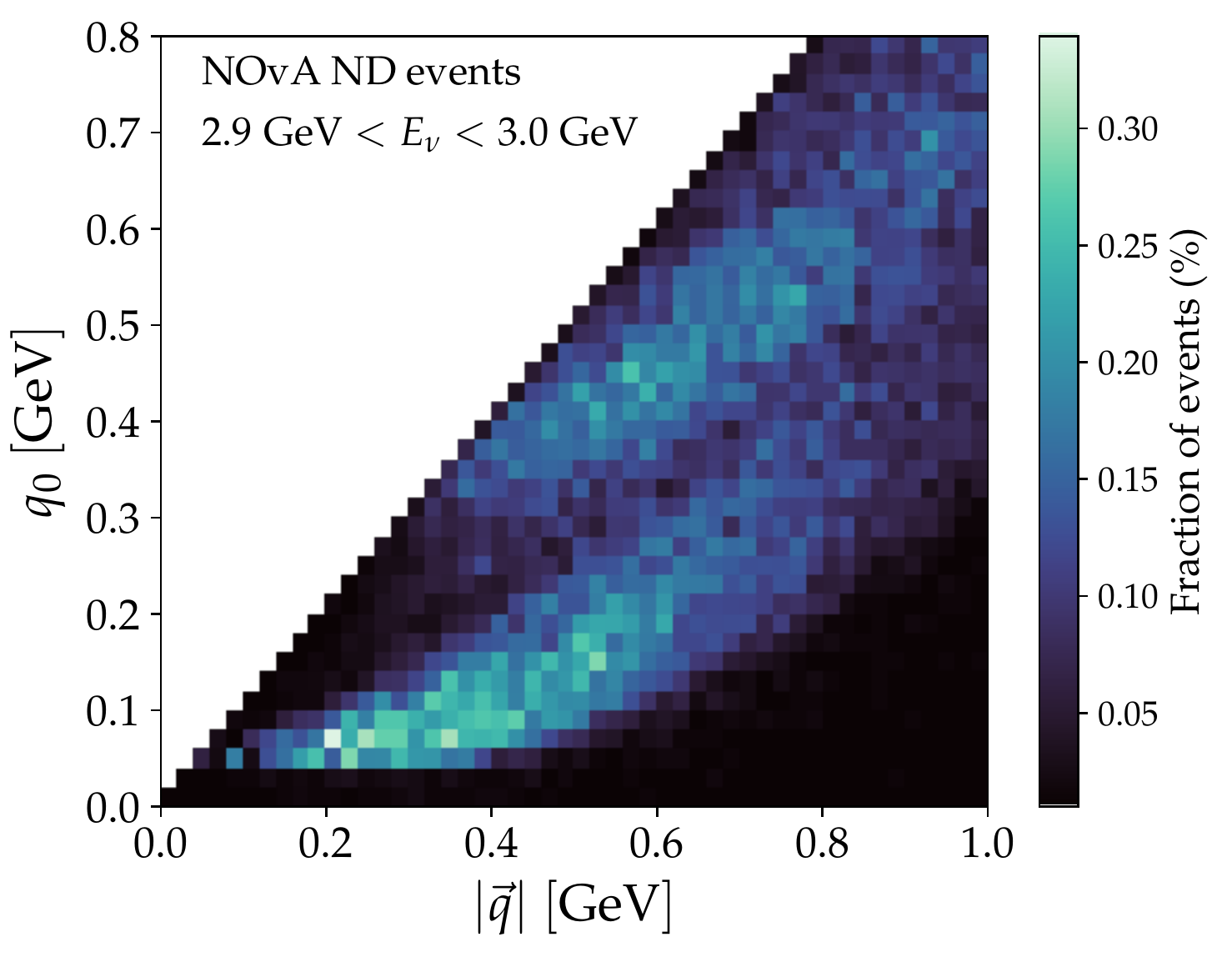}
	\caption{Normalized distribution of charged current $\nu_\mu$-carbon events at NOvA for $E_\nu\in[0.5,\,0.6]$~GeV (left panel) and $E_\nu\in[2.9,\,3]$~GeV (right panel).
	The large overlap for these two disparate energies shows the lack of correlation between neutrino energy and NOvA's tuning variables.
	\label{Fig:sterile_tune1}}
\end{figure}

In Sec.~\ref{Sec:ResultsSterile} we have shown that the ND tuning essentially does not change the shape of the reconstructed neutrino energy, and therefore it does not wash out the spectral wiggles coming from short baseline oscillations. 
The main reason for this is the lack of a clear correlation between the tuning parameters $(|\vec{q}\,|,\,q^0)$ and the neutrino energy.
We can appreciate this in Fig.~\ref{Fig:sterile_tune1}, where we present the normalized distribution of events in the NOvA near detector, restricting the true neutrino energy to two disparate regions, $E_\nu\in[0.5,\,0.6]$~GeV (left panel) and $E_\nu\in[2.9,\,3.0]$~GeV (right panel).

While there are visible differences in these distributions, the large overlap between them shows the lack of correlation among tuning variables and neutrino energy.
The main goal of the experiment is to measure the true neutrino energy spectra in both ND and FD so as to extract the oscillation probability. 
Therefore, the tuning is performed in parameters that are not fully correlated with neutrino energy. 
Regardless, while the lack of correlation prevents the tuning of washing out ND oscillations, it also makes the tuning unable to fix arbitrary cross section mis-modelings, which as discussed in Sec.~\ref{Sec:ResultsSterile} can bias the experimental sensitivity to sterile neutrinos.

\subsection{Light neutrinophilic scalars}
\label{subsec:mononu}

In Sec.~\ref{sec:Mononu}, we have seen a large discrepancy between the missing transverse momentum spectra generated by \texttt{GENIE} and \texttt{NuWro}, particularly after experimental cuts.
The goal of this section is to understand the origin of this discrepancy.
Let us first recall that we select events in antineutrino beam mode with exactly one proton above 100 MeV, no neutrons above 100 MeV and no pions.
Naively, only a small fraction of MEC events, in which a low energy neutron and a proton are emitted, contributes to this sample.
Nevertheless, final-state interactions will give rise to contributions to this sample coming from other cross section processes.
For example, although quasi-elastic scattering of antineutrinos on nucleons produce a final-state neutron, $\overline\nu_\mu+p\to\mu^+ + n$, as this neutron propagates throughout the nuclear medium it can kick out a proton, leading to an event that passes experimental cuts.
Even if at the NOvA energy scale these effects are subleading, we call attention to the fact that cuts imposed here suppress standard model backgrounds by two orders of magnitude, see Fig.~\ref{fig:mononu}.
Therefore, the experimental search is susceptible to percent-level differences between generators.

\begin{figure}[t!]
	\centering
		\includegraphics[width=0.75\textwidth]{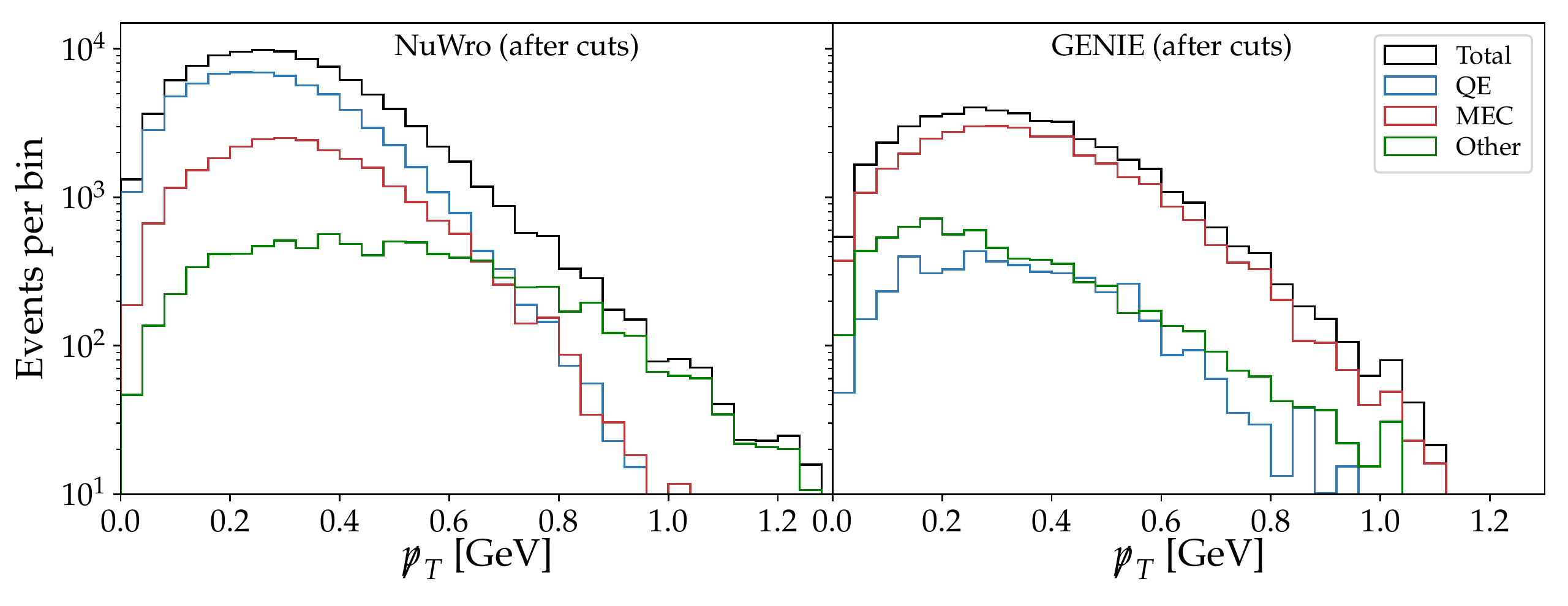}
	\caption{Missing transverse momentum spectrum broken down by neutrino-nucleus cross section process for \texttt{NuWro} (left) and \texttt{GENIE} (right), after experimental cuts. The cuts select events with exactly one proton above 100 MeV, no neutrons above 100 MeV and no pions.
	\label{fig:cut-processes}}
\end{figure}

In Fig.~\ref{fig:cut-processes}, we show the missing transverse momentum spectra broken down by neutrino-nucleus cross section process for \texttt{NuWro} (left) and \texttt{GENIE} (right), after experimental cuts. 
We can clearly see that the two generators strongly disagree on the physics behind the events that pass selection: while in \texttt{NuWro} the spectrum is dominated by quasi-elastic interactions (blue), in \texttt{GENIE} the MEC events (red) are much more prominent.
Since the magnitude of the quasi-elastic component of the cross section in the two generators are quite similar, this difference can only come from final-state interactions.
This already puts in question the focus on MEC-only tuning.
More importantly, given this wildly different physics in selected events between generators, there is no reason to expect that an unphysical tuning would mitigate the observed discrepancy.

\bibliographystyle{JHEP}
\bibliography{refs}

\end{document}